\documentclass[oldversion,preprint]{aa}
\usepackage{graphicx}
\usepackage{txfonts}
\begin{document}
   \title{A seasonal cycle and an abrupt change in the variability characteristics of the
     intraday variable source S4 0954+65}

   \author{N.~Marchili \inst{1,2} \and
           T.~P.~Krichbaum \inst{1} \and
           X.~Liu \inst{3} \and
           H.-G.~Song \inst{3} \and
           K.~\'E.~Gab\'anyi \inst{4,5} \and
	   L.~Fuhrmann\inst{1} \and
           A.~Witzel \inst{1} \and
           J.~A.~Zensus \inst{1}
          }

   \offprints{N. Marchili}

   \institute{Max-Planck-Institut f\"ur Radioastronomie, Auf dem H\"ugel 69,
              53121 Bonn, Germany\\
              \email{marchili@mpifr-bonn.mpg.de}
         \and
             Dipartimento di Astronomia, Universit\`a di Padova, Vicolo dellÕOsservatorio 3, 35122, Padova, Italy	
         \and
             Xinjiang Astronomical Observatory, the Chinese Academy of Sciences, 
             150 Science 1-Street, Urumqi 830011, China
         \and
             F\"OMI, Satellite Geodetic Observatory, Budapest, Hungary
         \and
	      Konkoly Thege Mikl\'os Astronomical Institute of the Research Centre
	for Astronomy and Earth Sciences, Hungarian Academy of Sciences
             }

   \date{Received 7 February 2012; Accepted 13 April 2012}

\abstract{The BLLac object S4~0954+65 is one of the main targets of the Urumqi monitoring program targeting IntraDay Variable (IDV) sources.
Between August 2005 and December 2009, the source was included in 41 observing sessions, carried out at a frequency of 4.8\,GHz. The time analysis of the collected light curves, performed by applying both a structure function analysis and a specifically developed wavelet-based algorithm, discovered an annual cycle in the variability timescales, suggesting that there is a fundamental contribution by interstellar scintillation to the IDV pattern of the source. The combined use of the two analysis methods also revealed that there was a dramatic change in the variability characteristics of the source between February and March 2008, at the starting time of a strong outburst phase. The analysis' results suggest that the flaring state of the source coincides with the appearance of multiple timescales in its light curves, indicating that changes in the structure of the relativistically moving emitting region may strongly influence the variability observed on IDV timescales.}


\authorrunning{N. Marchili et al.}
\titlerunning{The variability characteristics of the
     IDV source S4~0954+65}

   \maketitle
%
\section{Introduction}

Radio variability on timescales from days to weeks was observed in AGNs as
early as 1971 (see, e.g., Kinman \& Conklin \cite{Kinman1971}, Wills \cite{Wills1971}). 
In 1986, Witzel et al. (\cite{Witzel1986}) reported the discovery of flux density
variations occurring on timescales shorter than one day, which have
been referred to as IntraDay Variability (IDV). It is known
that IDV affects 20\% to 50\% of the flat-spectrum radio source
population (see Quirrenbach et al. \cite{Quirrenbach1992}, Lovell et
al. \cite{Lovell2008}). The origin of the variability, however, is not
completely understood. Assuming that the variability is intrinsic to
the source, causality arguments constrain the maximum
size of the emission region using the observed variability timescales. For 
IDV sources, this frequently implies very high brightness
temperatures, exceeding by far the inverse-Compton limit ($10^{12}$\,K;
see Kellermann \& Pauliny-Toth \cite{Kellermann1969}).   

Propagation effects, such as InterStellar Scintillation (ISS), were suggested
as a likely alternative scenario. In this model, the variability is caused by the
scintillation of very compact components (of the size of tens of $\mathrm{\mu}$as) in
the IDV source by a scattering screen placed between the source and the 
observer. The radiation can be either focused or defocused by the screen, 
causing the measured flux density to be a function of the line of sight 
to the observer. Owing to the Earth's revolution
around the Sun, the observer's line of sight towards the screen changes 
continuously, causing the detected rapid variations in the flux density 
measurements. Consequently, the faster the Earth's movement with respect 
to the scattering screen, the shorter the variability timescale. 
Since the Earth's velocity follows a one-year periodic cycle, the relative 
velocity between the scattering screen and the observer should change 
accordingly, resulting in a seasonal cycle of the variability timescale. 

The detection of annual modulation patterns has been essential in 
demonstrating the source-extrinsic origin of IDV for the most extreme 
sources (timescales of hours, amplitudes of $\sim$100\%), the 
so-called fast scintillators, namely PKS~0405-385, J1819+384, and
PKS~1257-326 (see Kedziora-Chudczer et al. \cite{Kedziora1997},
Dennett-Thorpe \& de Bruyn \cite{Dennett2000}, Bignall et al. \cite{Bignall2003}). 
For type-II IDV sources (characterized by less extreme variability, with amplitudes 
of $\sim1-20$\% and timescales $\lesssim$ 2 days, see, e.g., Quirrenbach et al. 
\cite{Quirrenbach1992}), the situation is less clear. It may partially depend
on their relatively long variability timescales, which make seasonal cycles harder to 
identify, given the unrealistically long observing time that would be required 
to measure accurate timescale estimates. Seasonal cycles have been detected in J1128+592 (see 
Gab\'anyi et al. \cite{Gabanyi2007}) and claimed for other sources, such as 
S4~0917+62 (see Rickett et al. \cite{Rickett2001}). The MASIV survey (Lovell 
et al. \cite{Lovell2003} and \cite{Lovell2008}) showed that a significant part of the 
flux density variations observed in compact radio sources are due to ISS. On the 
other hand, the discovery of correlated optical-radio IDV in the flux density of the 
type-II source S5~0716+71 suggested that its origin may be intrinsic to the source 
(see Quirrenbach et al. \cite{Quirrenbach1991}).

The detailed analysis of the variability characteristics of IDV sources
plays an essential role in understanding the ISS contribution to their variability
pattern. From 2005 to the end of 2009, the Max-Planck-Institut f\"ur
Radioastronomie and the Urumqi Observatory carried out a program of
intensive monitoring of type-II IDV sources with the Nanshan 25m
radio telescope, located about 70\,km south of Urumqi city (China). The main
aims of the project are: (i) monitoring the long-term IDV characteristics of a larger 
sample of IDV sources, looking for seasonal cycles in their variability patterns,
and (ii) detecting new IDV sources. The main target of the
monitoring program were the type-II sources AO~0235+164, S5~0716+71, 
S4~0917+62, S4~0954+65, J1128+592, and HB89~1156+295. During the 
observations, about sixty additional sources were sporadically observed 
while attempting to find new IDV candidates. 

In the present paper, we focus on our results obtained for S4~0954+65, which is a 
well-studied BLLac object at redshift $z=0.368$ whose IDV features have been investigated
since the early '90s (Wagner et al. \cite{Wagner1993}). The origin of the variability
is still controversial --- S4~0954+65 is the second source for which a radio-optical correlation 
has been claimed (Wagner et al. \cite{Wagner1990}). However, the detection of 
extreme scattering events (Fiedler et al. \cite{Fiedler1987}, \cite{Fiedler1994}; 
Cim\`o et al. \cite{Cimo2002}) and the possible presence of a seasonal cycle in
the variability timescales of the source, which was revealed by a previous IDV
monitoring project (see Fuhrmann \cite{Fuhrmann2004}), seem to indicate that 
at least part of the variability is caused by extrinsic mechanisms, and that more than
one scattering screen may be interposed between the source and the Earth.

In the next section, we present an overview of the observations
and the data calibration procedure. The analysis methods used to 
extract the variability characteristics of the light curves are
presented in Sec.\,3. The main results are shown in Sec.\,4. In Sec.\,5 
and 6, we discuss our results and present a summary of the paper, respectively. 

\section{Observation and data calibration}

The Nanshan 25m telescope is equipped with a 4.8\,GHz single-horn dual-polarization
receiver provided, along with the new telescope driving program, by the MPIfR
(for more details, see Sun et al. \cite{Sun2006}). The receiver's bandwidth is 600\,MHz. The
flux density measurements were performed in cross-scan mode. Each scan
consists of eight sub-scans in perpendicular directions across the source position
in order to estimate the pointing offsets in the two scanning
directions and correct for them. The raw data were processed at the MPIfR via
TOOLBOX, a Python-based software package that
subtracts baseline drifts and fits a Gaussian profile to each cross-scan. The
amplitude of the profile provides an estimate of the source's flux density
expressed in units of antenna temperature (K).

The data calibration was performed in semi-automatic mode (for more
details, see Marchili \cite{MarchiliTh}, Marchili et al. \cite{Marchili2010}). The quality 
of the individual sub-scans was 
checked by considering their pointing offset, the full-width at half-maximum (FWHM) 
of their Gaussian profile, and the error in their flux density measurement. 
Sub-scans that showed pointing offsets of more than $100''$, deviations from 
the average FWHM (i.e. $\sim 600''$) larger than $150''$, or uncertainties in 
the Gaussian fit estimate larger than 10\% of the amplitude were automatically 
discarded. Subsequently, a weighted average of the remaining sub-scans was 
obtained with a weight proportional to the inverse of
the squared errors, separately for the two scanning directions. After the
pointing-error correction, we averaged the mean sub-scan amplitudes in the two scanning directions, providing a single flux-density measurement per
scan. 

In each observing session, about $50\%$ of the time was dedicated to 
measurements of primary and secondary calibrators, i.e., compact radio-sources 
with steep spectra and no evidence of IDV. The 5\,GHz flux densities of the 
so-called primary calibrators are fairly constant in time and well-known. The 
secondary calibrators, instead, show variability on timescales of
months. We used the primary and secondary calibrators to estimate the
variations in the gain of the antenna due to changes in both its parabolic
shape and the position of the focus. We used the secondary calibrators --- 
which were selected to be at small
angular separations from the target sources --- to correct for possible changes
in the collected flux densities caused by instabilities in the
antenna-receiver system or changing weather conditions. The data calibration was completed by the conversion of measured antenna temperatures (in K) to
flux densities (in Jy). The time-interpolated conversion factors were derived by means of the known flux densities of the primary calibrators, namely NGC~7027, 3C~286, and 3C~48. The complete data-calibration
procedure guarantees a high level of accuracy, which, based on the
residual variability in the calibrators, can be estimated in the range
between 0.2 and 0.8\%, under normal weather conditions.

\begin{figure}[tbp!]
   \centering
   \includegraphics[width=0.96\columnwidth]{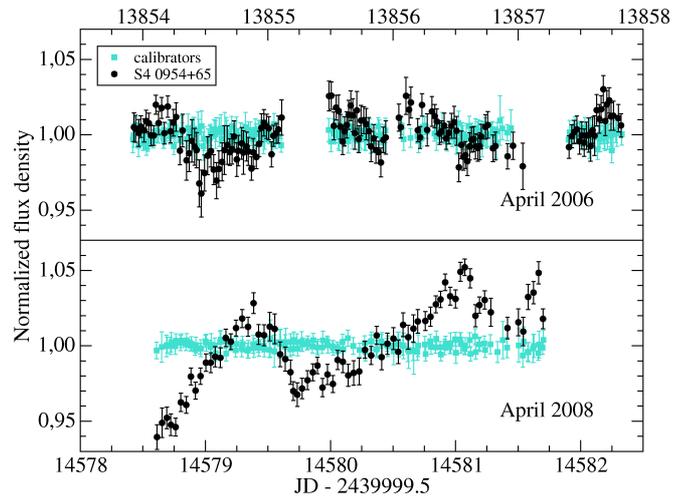}
      \caption{The variability curves of S4~0954+65 (black dots) in April 2006 (upper panel)
      and April 2008 (lower panel). The turquoise squares show the residual variability in
      the fluxes of the secondary calibrators S5~0836+71 and S4~0951+69.
              }
              \label{fig:lc}
\end{figure}

During the monitoring campaign, between August 2005 and December 2009, 
41 IDV datasets for S4~0954+65 were collected. Two examples of light curves
are shown in Fig. \ref{fig:lc}. In three cases, the observing sessions were too short 
($\le 1.7$\,d) to provide a reasonable estimation of the variability characteristics 
of the sources, were consequently discarded from the analysis we describe 
in the following. An overview 
of the main characteristics of the remaining 38 sessions is reported in Table
\ref{tab:epochs}. In the few cases in which the duration of the observations is 
longer than 5 days or shorter than 2.5 days, a bias of the estimated timescales 
toward, respectively, the slowest and the fastest variability components cannot be 
excluded.

\begin{table}[h]
\caption{Basic information about the 38 observing sessions at the Urumqi 
	Observatory during which S4~0954+65 was observed. In Col. 1, we 
	report the starting and ending dates of the experiments (format: 
	dd.mm$-$dd.mm.yyyy), in Col. 2 the duration, and in Col. 3 the average 
	number of measurements per hour for  IDV sources (duty cycle).
	\label{tab:epochs}}  
\centering
\begin{tabular}{c c c}
\hline
\hline
\noalign{\smallskip}
 Epoch  & Duration   & Duty cycle \\
        &    (d)     & (h$^{-1}$) \\
\noalign{\smallskip}
\hline
\noalign{\smallskip}
14.08$-$17.08.2005   &    2.9   & 0.3 \\
27.12$-$31.12.2005   &    3.7   & 0.6 \\
15.03$-$18.03.2006   &    3.0   & 0.7 \\
27.04$-$01.05.2006   &    3.9   & 0.7 \\
09.06$-$12.06.2006   &    3.2   & 1.2 \\
14.07$-$18.07.2006   &    4.0   & 0.9 \\
19.08$-$25.08.2006   &    6.4   & 1.3 \\
23.09$-$28.09.2006   &    5.0   & 1.2 \\
17.11$-$22.11.2006   &    4.7   & 1.2 \\
17.12$-$21.12.2006   &    4.2   & 1.0 \\
25.01$-$27.01.2007   &    2.3   & 1.1 \\
12.02$-$16.02.2007   &    4.0   & 1.0 \\
24.03$-$27.03.2007   &    2.8   & 0.9 \\
20.04$-$24.04.2007   &    3.7   & 0.8 \\
15.06$-$18.06.2007   &    2.4   & 0.9 \\
19.07$-$22.07.2007   &    2.9   & 0.9 \\
18.08$-$21.08.2007   &    3.1   & 0.9 \\
13.10$-$16.10.2007   &    3.0   & 1.0 \\
21.12$-$25.12.2007   &    3.2   & 1.1 \\
24.02$-$27.02.2008   &    2.9   & 0.8 \\
21.03$-$24.03.2008   &    3.0   & 1.1 \\
21.04$-$24.04.2008   &    3.1   & 1.0 \\
21.06$-$24.06.2008   &    3.5   & 0.6 \\
18.07$-$23.07.2008   &    4.7   & 0.6 \\
20.08$-$25.08.2008   &    5.0   & 0.6 \\
12.09$-$16.09.2008   &    3.5   & 1.0 \\
06.11$-$09.11.2008   &    3.6   & 0.5 \\
22.12$-$24.12.2008   &    2.4   & 1.0 \\
11.01$-$13.01.2009   &    2.6   & 1.1 \\
22.02$-$25.02.2009   &    3.0   & 1.4 \\
19.04$-$24.04.2009   &    5.5   & 0.7 \\
12.05$-$15.05.2009   &    2.7   & 0.7 \\
24.06$-$27.06.2009   &    2.6   & 0.7 \\
20.08$-$24.08.2009   &    4.1   & 0.9 \\
22.09$-$27.09.2009   &    5.5   & 0.9 \\
08.10$-$11.10.2009   &    2.3   & 1.0 \\ 
21.11$-$25.11.2009   &    3.8   & 0.8 \\
10.12$-$14.12.2009   &    4.4   & 1.1 \\
\hline
\end{tabular}
\end{table}

\section{Variability analysis}

Following the scheme proposed by Quirrenbach et
al. (\cite{Quirrenbach2000}) and Kraus et al. (\cite{Kraus2003}), two 
different quantities were used to evaluate the significance and
amplitude of the variability, namely the reduced chi-squared test statistic
${\chi_{\mathrm{r}}}^2$ and the variability amplitude $Y$. 

The reduced chi-square (see, e.g.,  Bevington \& Robinson \cite{Bevington1992})
was used to test for given datasets the likelihood of the hypothesis of a constant flux density

\begin{equation}
{\chi_{\mathrm{r}}}^2=\frac{1}{N-1}
\sum_{i=1}^N{\Big(\frac{S_{\mathrm{i}}-<S>}{\delta S_{\mathrm{i}}}\Big)}^2,
\end{equation}
where N is the number of datapoints, $<~S~>$ the average flux density,
$S_{\mathrm{i}}$ the i-th flux density measurement, and $\delta S_{\mathrm{i}}$
its error. A dataset was regarded as variable if the ${\chi_{\mathrm{r}}}^2$ 
test returned a probability of constant flux below 0.1\%, which is equivalent 
to a 99.9\% probability for significant variability.

An estimation of the variability in a dataset was provided by the modulation index, 
$m_{\mathrm{i}}[\%]=100~{\sigma_{\mathrm{S}}}/{<S>}$, where $\sigma_{\mathrm{S}}$ 
is the root mean square (rms) in flux density. However, this parameter does not take into account 
the residual noise in the calibrated data, which varies from epoch to epoch, mainly 
depending on weather conditions. We defined the variability amplitude Y as an equivalent to a 3-$\sigma$ confidence interval for Gaussian noise,
modified by the quadratic subtraction of the rms noise floor, $m_{\mathrm{0}}$, to take into account the noise bias

\begin{equation}
Y=3\sqrt{m_{\mathrm{i}}^2-m_{\mathrm{0}}^2}.
\label{eq:y}
\end{equation}

\subsection{Variability timescale}

The characteristic timescales of the variable datasets were estimated
using two different methods of time series analysis --- a
first-order structure function (see Simonetti et al. \cite{Simonetti1985}) 
and a wavelet-based analysis.

\begin{figure}[tbp]
   \centering
   \includegraphics[width=0.96\columnwidth]{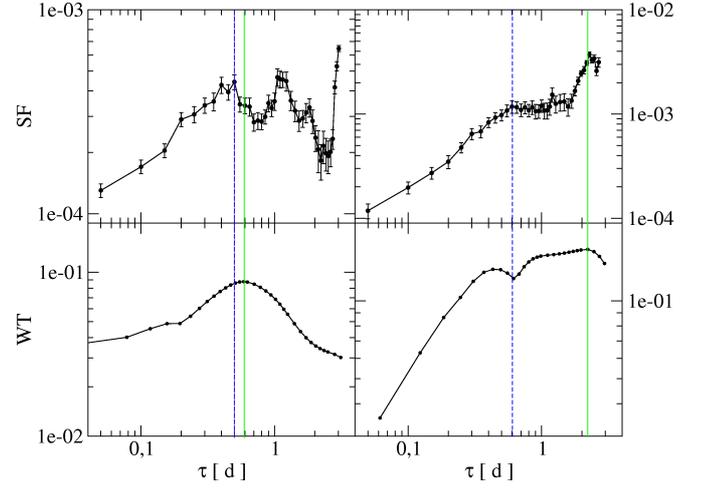}
      \caption{Upper panels: the structure function plots of the April 2006 
      (left panel) and April 2008 (right panel) datasets; the blue vertical 
      lines show the estimated timescales. Lower panels: the wavelet-based 
      analysis for the same epochs; the green vertical lines show the 
      estimated timescales. The existence of two bumps in the right
      panels is the signature of two distinct variability components in the dataset.
              }
              \label{fig:tam}
\end{figure}

The structure function of a given time series $\{S(t_i)\}_i$ is calculated as

\begin{equation}
SF(\tau)=\frac{1}{n}\sum_{ij}[S(t_i)-S(t_j)]^2,
\end{equation}
where the sum is extended to the n pairs $(t_i, t_j)$ for which
$\parallel~t_i-t_j~\parallel \simeq \tau$. Ideally, when a structure function
is applied to a red-noise-like signal, it shows a monotonic increase that
can be described as a power law. At the time lag $\tau_\mathrm{sf}$ corresponding
to the maximum coherency time of the signal, the structure function shows 
a plateau, for which $\tau_\mathrm{sf}$ provides an estimate of the characteristic 
timescale of the signal. Owing to both the uneven sampling and the noise in 
the analyzed data, the position of the plateau is subject to some degree of 
uncertainty. For each plateau, it is possible to determine a minimum and a 
maximum timescale --- respectively $\tau_{\mathrm{min}}$ and $\tau_{\mathrm{max}}$. 
The characteristic timescale of the structure function $\tau_{\mathrm{sf}}$ was
calculated to be $(\tau_{\mathrm{max}}+\tau_{\mathrm{min}})/2$.
The uncertainty in the timescale estimation was evaluated as the 
maximum between two terms, namely $(\tau_{\mathrm{max}}-\tau_{\mathrm{min}})/2$ 
(which reflects the significance of the plateau) and 
$0.35\cdot ({\tau_{\mathrm{sf}}}^{3/2})/{(obs_d)}^{1/2}$ (see Marchili 
et al. \cite{Marchili2011}), where $obs_d$ indicates the duration of the observation. The latter term accounts for the significance of the timescale, which is limited 
by the small number 
of cycles  that it is possible to observe, given the limited duration of the observations.
To estimate the proportionality factor, we calculated all the time-intervals
between a local minimum (or maximum) and the following maximum (or minimum) for a set of light curves characterized by rapid variability; for each light curve, we
calculated the average of the observed time-intervals (which provides an 
estimate of the timescale that is consistent with ${\tau_{\mathrm{sf}}}$) and 
their standard deviation (which provides an estimate of its uncertainty). A linear
regression of the uncertainties, plotted against the values of
(${\tau_{\mathrm{sf}}}^{3/2})/{(obs_d)}^{1/2}$ for the corresponding light curves, 
returned the proportionality factor 0.35.

The structure function in some cases was found to have more than one plateau,
which is indicative of multiple variability timescales. In these cases,
we identified the characteristic timescale with the shortest one. 

The wavelet-based algorithm that we developed is based on the Ricker
(`Mexican hat') mother wavelet

\begin{equation}
\Psi(t)=C\,\big(1-t^2\big)\,exp\big(-\frac{t^2}{2}\big)
\end{equation}
where $C$ is a normalization factor and $t$ is the time parameter. An extensive description of this analysis method can be found in Appendix A. The timescales
obtained by applying the wavelet analysis are indicated with $\tau_\mathrm{wt}$.

Examples of structure functions and wavelet results are provided in Fig. \ref{fig:tam} 
for the light curves shown in Fig. \ref{fig:lc}.

\section{Results}
\label{Results}

The variability characteristics of S4~0954+65 are summarized in Table
\ref{tab:res}. Our study had
two main outcomes, which are discussed below: (i) we found 
strong evidence of an abrupt change in the variability characteristics of
S4~0954+65 between February and March 2008; (ii) the IDV timescales of the source 
seem to undergo an annual modulation.

\begin{table}[h]
\caption{The variability characteristics of S4~0954+65 in 38 epochs of
  observation with the Urumqi Telescope.  For each session (Col. 1), 
  we report the average flux density (Col. 2), the variability amplitude (Col. 3), 
  the reduced chi-square (Col. 4), the wavelet and structure function timescales 
  (Col. 5 and 6, respectively), and the number of data points in each light curve 
  (Col. 7).\label{tab:res}}  
\centering
\begin{tabular}{l c c r @{.} l c c r}
\hline
\hline
\noalign{\smallskip}
 Epoch  & S$_\mathrm{5GHz}$ & $Y$ & \multicolumn{2}{c}{${\chi}_\mathrm{r}^2$}  & $\tau_\mathrm{wt}$ &  $\tau_\mathrm{sf}$ & data \\
        &  [Jy]  &  [\%] & \multicolumn{2}{c}{} & [d] & [d] & \\
\hline
\noalign{\smallskip}
\hline
\noalign{\smallskip}
2005.08.14 &  0.92 &  5.08  &   3&47 &  0.9$\pm$0.1 &  1.2$\pm$0.3  &   46 \\
2005.12.27 &  1.14 &  5.29  &   4&39 &  2.6$\pm$0.6 &  3.0$\pm$0.9  &   54 \\
2006.03.15 &  0.92 &  4.54  &   3&13 &  0.6$\pm$0.1 &  0.6$\pm$0.1  &   50 \\
2006.04.27 &  1.14 &  3.44  &   1&65 &  0.6$\pm$0.1 &  0.5$\pm$0.1  &  145 \\
2006.06.09 &  1.11 &  5.23  &   4&72 &  2.4$\pm$0.5 &  0.9$\pm$0.2  &   86 \\
2006.07.14 &  1.02 &  2.35  &   1&08 &   ---                &  ---                   &   81 \\
2006.08.19 &  1.09 &  2.63  &   1&49 &  1.3$\pm$0.3 &  1.4$\pm$0.3  &  169 \\
2006.09.23 &  1.21 &  3.22  &   1&81 &  0.7$\pm$0.1 &  0.8$\pm$0.2  &  134 \\
2006.11.17 &  1.07 &  2.83  &   1&79 &  2.4$\pm$0.7 &  4.0$\pm$1.3  &  134 \\
2006.12.17 &  0.96 &  1.76  &   1&56 &  0.4$\pm$0.1 &  0.7$\pm$0.1  &   97 \\
2007.01.25 &  1.08 &  2.60  &   1&89 &  0.4$\pm$0.1 &  0.6$\pm$0.1  &   66 \\
2007.02.12 &  1.06 &  6.71  &   7&99 &  1.3$\pm$0.6 &  1.8$\pm$0.5  &  109 \\
2007.03.24 &  1.29 &  3.17  &   3&05 &  0.5$\pm$0.1 &  0.4$\pm$0.1  &   70 \\
2007.04.20 &  1.30 &  4.80  &   4&74 &  1.7$\pm$0.4 &  1.6$\pm$0.4  &   65 \\
2007.06.16 &  1.02 &  3.64  &   2&01 &  0.3$\pm$0.1 &  0.5$\pm$0.1  &   57 \\
2007.07.19 &  0.90 &  4.90  &   3&05 &  0.6$\pm$0.2 &  0.7$\pm$0.2  &   62 \\
2007.08.18 &  0.86 &  5.41  &   3&62 &  0.3$\pm$0.1 &  0.3$\pm$0.1  &   73 \\
2007.10.13 &  0.94 &  2.93  &   2&81 &  0.9$\pm$0.4 &  0.6$\pm$0.1  &   60 \\
2007.12.22 &  1.13 &  2.71  &   3&09 &  0.3$\pm$0.1 &  0.5$\pm$0.1  &   85 \\
2008.02.25 &  0.95 &  3.91  &   4&91 &  0.3$\pm$0.1 &  0.3$\pm$0.1  &   54 \\
\hline
2008.03.22 &  0.90 &  6.60  &  11&45 & 1.1$\pm$0.4 &  0.4$\pm$0.2  &   74 \\
2008.04.22 &  0.98 &  7.73  &  16&45 & 2.2$\pm$0.6 &  0.6$\pm$0.1  &   73 \\
2008.06.21 &  1.23 &  5.32  &   9&56 &  0.7$\pm$0.1 &  0.7$\pm$0.1  &   57 \\
2008.07.18 &  1.32 &  6.40  &  14&88 & 2.5$\pm$0.2 &  0.3$\pm$0.1  &   72 \\
2008.08.20 &  1.33 &  3.90  &   6&10 &  1.1$\pm$0.3 &  1.0$\pm$0.3  &   75 \\
2008.09.12 &  1.39 &  1.73  &   2&10 &  1.0$\pm$0.5 &  0.4$\pm$0.2  &   88 \\
2008.11.06 &  1.55 &  4.61  &   8&34 &  2.3$\pm$0.5 &  0.7$\pm$0.1  &   52 \\
2008.12.22 &  1.27 &  3.37  &   5&26 &  1.0$\pm$0.4 &  0.3$\pm$0.1  &   59 \\
2009.01.11 &  1.11 &  2.15  &   2&30 &  1.3$\pm$0.3 &  0.2$\pm$0.1  &   65 \\
2009.02.23 &  0.95 &  3.65  &   3&65 &  0.9$\pm$0.2 &  1.0$\pm$0.3  &   76 \\
2009.04.19 &  0.92 &  3.99  &   4&09 &  0.5$\pm$0.1 &  0.6$\pm$0.1  &   88 \\
2009.05.13 &  0.96 &  4.42  &   5&86 &  2.2$\pm$0.4 &  1.0$\pm$0.3  &   44 \\
2009.06.26 &  1.01 &  3.39  &   3&05 &  1.0$\pm$0.2 &  0.4$\pm$0.2  &   43 \\
2009.08.21 &  1.07 &  6.00  &   9&57 &  0.9$\pm$0.1 &  1.3$\pm$0.3  &   87 \\
2009.09.22 &  1.22 &  6.10  &  11&02 & 0.9$\pm$0.1 &  1.5$\pm$0.4  &  121 \\
2009.10.09 &  1.22 &  1.60  &   1&38 &   ---                &  ---                  &   56 \\
2009.11.22 &  1.27 &  4.97  &   5&38 &  0.8$\pm$0.1 &  1.0$\pm$0.2  &   65 \\
2009.12.11 &  1.27 &  4.44  &   6&51 &  1.7$\pm$0.4 &  0.5$\pm$0.2  &  113 \\
\hline
\end{tabular}
\end{table}

\begin{figure}[tbp]
   \centering
   \includegraphics[width=0.96\columnwidth, height=0.8\columnwidth]{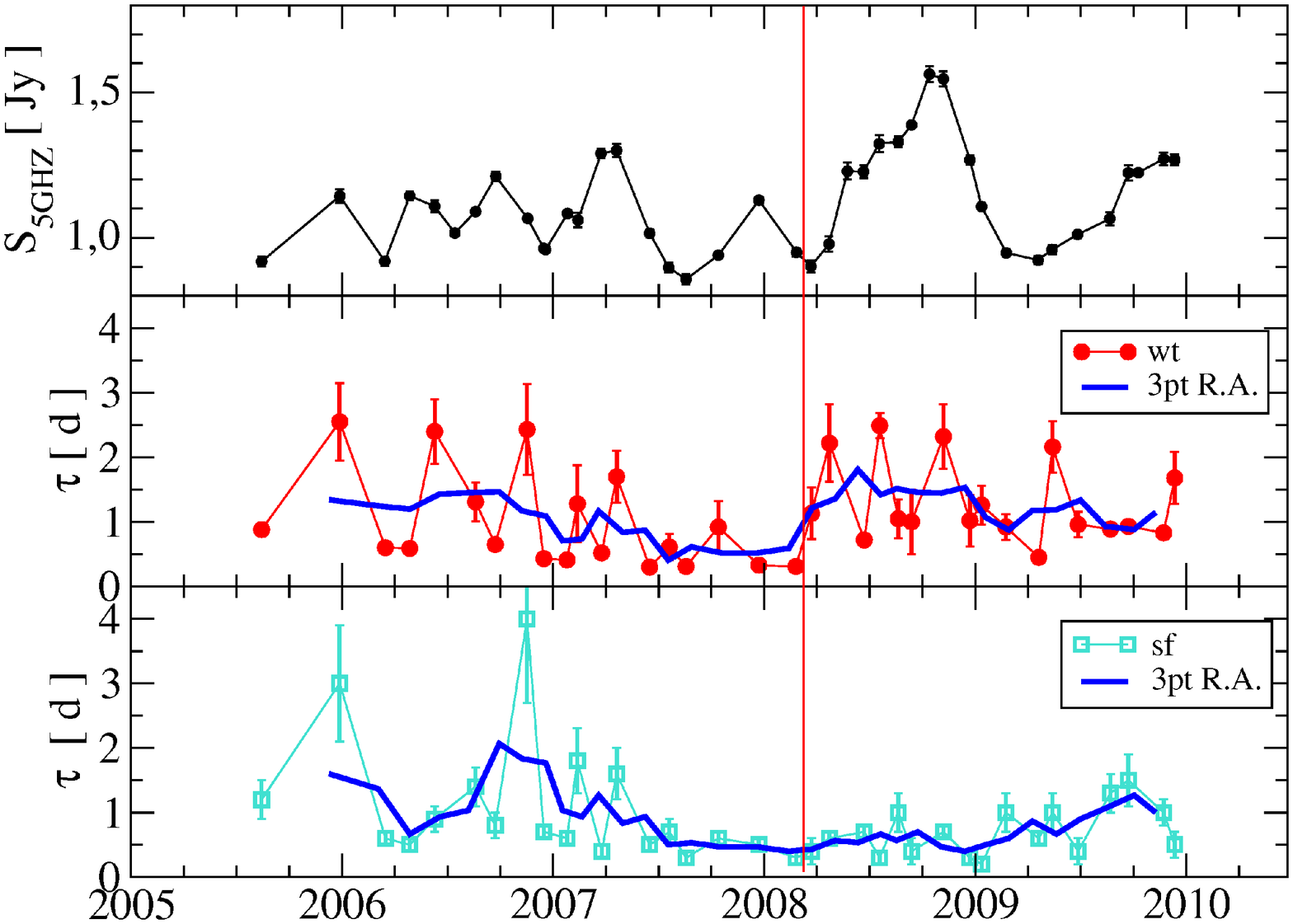}
      \caption{The long-term variability of S4~0954+65 (upper panel). In the middle panel,
      the wavelet timescales (red dots) and a three-point running average (blue line), showing
      the sudden increase starting from March 2008. The structure function timescales are
      shown in the lower panel. The red vertical line divides the observing sessions into two 
      phases characterized by remarkably different variability characteristics.
              }
              \label{fig:0954cng}
\end{figure}

\subsection{Change in the overall variability characteristics}

In the 
long-term light curve of the source (Fig. \ref{fig:0954cng}, upper panel), two considerably
different phases can be recognized. From August 2005 up to February 2008, the flux density
variations were generally moderate in intensity (the average peak-to-peak variation is $\sim0.2$\,Jy) 
and occurred on timescales of three months; March 2008 marked the starting point of a strong outburst 
phase, which up to December 2009 was still ongoing,
characterized by large variations ($\sim0.4$\,Jy) on a timescale of about seven months. For simplicity, 
we refer to the first and the second timespan, respectively, as the pre-outburst and the outburst 
phase of S4~0954+65.

\begin{figure}[tbp]
   \centering
   \includegraphics[width=0.96\columnwidth]{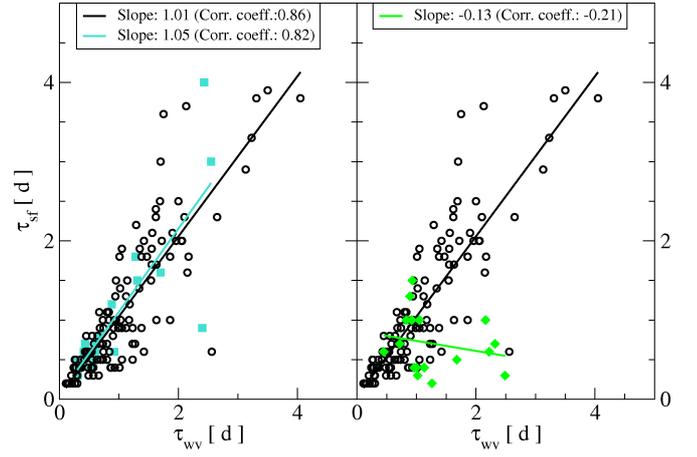}
      \caption{
      Comparison between the structure function and wavelet results for the comparison sample
      (black dots) and the S4~0954+65 light curves. In the left panel, the source's timescales 
      estimated up to February 2008 (cyan squares); in the right panel, the timescales from March
      2008 to December 2009 (green squares). The black, cyan, and green lines show the results 
      of applying linear regressions to the respective datasets.
              }
              \label{fig:ts_comp}
\end{figure}

\begin{figure}[tbp!]
   \centering
   \includegraphics[width=0.96\columnwidth]{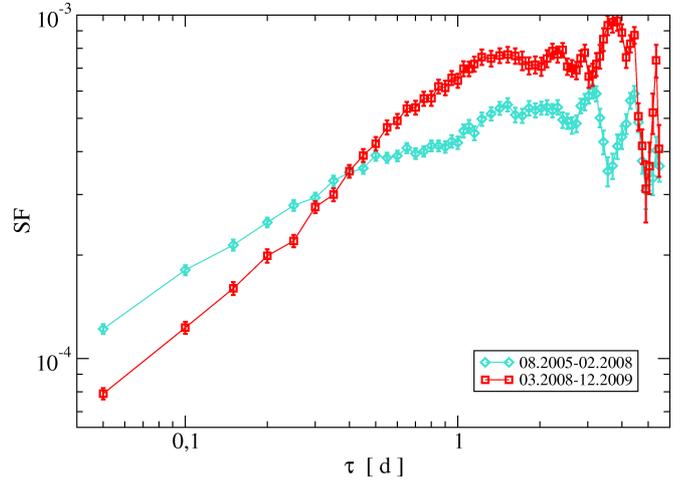}
      \caption{
      	The average structure function calculated from the normalized light curves 
      	obtained during the pre-outburst phase (turquoise diamonds) and the outburst phase
      	(red squares).           
      		}
              \label{fig:sfn}
\end{figure}

The change in the long-term activity of the source was also reflected in the 
remarkable variation in its IDV characteristics. There are three lines of evidence 
for such a variation:

\begin{itemize}
\item During the pre-outburst phase of the source, the IDV timescales provided by
the wavelet-based method and the structure function (see Fig. \ref{fig:0954cng}, middle and lower panels) were highly correlated with each other. A linear regression 
applied to the timescales (Fig. \ref{fig:ts_comp}, left panel) results in a correlation 
coefficient of 0.82 and a slope of 1.05, which proves the consistency in the 
definition of timescales used for the two methods. During the outburst phase, a 
similar analysis provides a correlation coefficient of -0.11 with a slope of -0.19 
(Fig. \ref{fig:ts_comp}, right panel), implying that there is a clear lack of correlation. 
It is useful to apply a linear regression to a comparison sample 
comprising all the timescales estimated for the other main sources in the Urumqi 
monitoring program (i.e., AO~0235+164, S5~0716+71, S4~0917+62, J1128+592, and 
HB89~1156+295). This returns a correlation coefficient of 0.86 with a slope of 1.01, 
which proves the high degree of compatibility between the structure function and the 
wavelet-based method results. This demonstrates that the variability 
characteristics of the source are atypical during the outburst phase, as illustrated by a 
Kolmogorov-Smirnov test for the comparison of datasets. There is a 74\% probability that 
the distribution of the values of $\tau_\mathrm{sf}/\tau_\mathrm{wt}$ during the pre-outburst 
phase is equal to the one estimated for the comparison sample. A similar test for the 
$\tau_\mathrm{sf}/\tau_\mathrm{wt}$ values during the outburst phase returns a probability 
below 0.1\%.
\item The wavelet timescales during the pre-outburst phase of S4~0954+65 follow a slowly
decreasing trend, pinpointed by the three-point running average in the middle panel of 
Fig.  \ref{fig:0954cng} (blue line). A sudden increase (from $\sim 0.5$ to $\sim 1.5$\,d) occurs in March 2008, after which the estimated timescales remain considerably longer than before. It is again helpful to 
apply a Kolmogorov-Smirnov test when comparing datasets; the probability that the 
two distributions of timescales are equal is below 1\%.
\item We calculated an average structure function from the light curves obtained during the 
pre-outburst phase, and compared it with the average one for the outburst phase\footnote{Note that, 
in the case of ISS-induced variability, the rms of the flux density is expected to 
increase linearly with the average flux density (see Narayan \cite{Narayan1992}). This implies that 
the structure functions calculated 
during the outburst phase are expected to be systematically higher than the ones obtained during 
the pre-outburst phase. To make the results comparable, we applied the structure function
to the light curves normalized by their average flux density.}. The difference 
between the two can be seen in Fig. \ref{fig:sfn}.
In the structure function of the pre-outburst phase, there is considerably more power on short timescales than in the structure function of the outburst phase. On longer timescales, we find the opposite situation, suggesting that there is some source of slower variability that is more dominant during the outburst phase.
\end{itemize}

Possible explanations of the large changes that occur in the IDV 
characteristics of S4~0954+65 at the beginning of its outburst phase, are 
discussed in section \ref{Discussion}. 

\subsection{Annual modulation of the timescales}

We investigated the possible annual cycle in the variability timescales of S4~0954+65
by analyzing the evolution of the timescale $\tau$ versus day of the year 
(DoY), separately for the wavelet-based and the structure function results. Following the 
scheme proposed by Qian \& Zhang (\cite{Qian2001}), updated to the case of 
anisotropic scattering (see, e.g., Bignall et al. \cite{Bignall2006}, Gab\'anyi et al. \cite{Gabanyi2007}), we expressed $\tau$ in terms of the vector ${\bf s}$ --- which defines the orientation of the 
elliptical scintillation pattern ---, the relative velocity between the scattering screen and the observer, 
${\bf v}$, the distance to the screen, D, and the anisotropy factor, r

\begin{equation}
\tau(\mathrm{DoY}) \propto \frac{\mathrm{D}\cdot \sqrt{\mathrm{r}}}{\sqrt{{\mathrm{v}^2(\mathrm{DoY})+(\mathrm{r}^2-1)\,({\bf v(\mathrm{DoY}) \times s})^2}}}.
\end{equation}
We note that the anisotropy may either be induced by an anisotropic scattering medium
or be intrinsic to the source (caused by an anisotropic emitting component).

We developed a code, based on a least squares fitting algorithm, to calculate 
the scintillation pattern parameters that most closely describe the detected variability timescales 
(for details, see Marchili \cite{MarchiliTh}). We note that the distance cannot be 
estimated without making `ad hoc' assumptions about the effective angular size 
of the source; in the following, we adopt the model proposed by Qian \& Zhang 
(\cite{Qian2001}), which assumes an intrinsic angular size of the scintillating 
component of 70 $\mu$as. In Section 5.2, we modify this size estimate, to obtain an improved fit to the detected variation in the IDV pattern.

\begin{table*}
\caption{The scintillation pattern parameters deduced by the distribution of the timescales
as a function of the day of the year. In Col. 1, we report the timespan 
taken into account for the estimation of the parameters; in Col. 2, the method used for 
estimating the timescales; in Col. 3 and 4, the relative velocity between the scattering screen
and the observer, projected into right ascension and declination; in Col. 5, the distance to 
the screen; in Col. 6 and 7, the anisotropy degree and angle.\label{tab:amp}}  
\centering
\begin{tabular}{c c c c c c c}
\hline
\hline
\noalign{\smallskip}
 Epochs & Method & v$_{\mathrm{ra}}$ & v$_{\mathrm{dec}}$ & D & r & $\theta$\\
              &              &   [km/s] & [km/s] & [kpc] & & [deg]\\
\hline
\noalign{\smallskip}
08.2005 -- 12.2009 &  WT & $10\pm3$ & $10\pm5$ & $0.27\pm0.03$ & $2.3\pm0.3$ & $84\pm6$\\
08.2005 -- 02.2008 &  WT & $11\pm5$ & $\ \ 8\pm7$ & $0.25\pm0.04$ & $3.0\pm0.5$ & $90\pm6$\\
03.2008 -- 12.2009 &  WT & $11\pm5$ & $11\pm5$ & $0.27\pm0.04$ & $1.7\pm0.3$ & $\ \ 82\pm12$\\
\hline
\noalign{\smallskip}
08.2005 -- 12.2009 &  SF  & $12\pm6$ & $-1\pm5$  & $0.18\pm0.03$ & $1.8\pm0.3$ & $\ \ 90\pm10$\\
08.2005 -- 02.2008 &  SF  & $12\pm5$ & $-5\pm6$  & $0.23\pm0.04$ & $1.7\pm0.3$ & $\ \ 82\pm12$\\
03.2008 -- 12.2009 &  SF  & $\ \ 5\pm9$ &  $\ \ 9\pm9$  & $0.14\pm0.08$ & $1.3\pm0.3$ & $\ \ 33\pm15$\\
\hline
\end{tabular}
\end{table*}

The plots of the timescales versus DoY --- the so-called annual modulation 
plots ---, comprising all the estimated timescales between August 2005 and December 
2009, are shown in Fig. \ref{fig:am1}, separately for the results provided by the two
analysis methods. The wavelet results (left panels in the figure) suggest 
that two slow-down phases occur during the year, centered on May (DoY 140) and 
November (DoY 312), corresponding to a cycle of approximately six months. The structure 
function timescales (right panels in the figure) show a less clear cycle, with only two 
data-points being considerably higher than the others, corresponding to the observing sessions of 
December 2005 and November 2006. Nevertheless, they appear to confirm that there is an annual cycle in the 
timescales that has a prominent slow-down phase peaking in November (DoY 320), 
and a much weaker one peaking in May (DoY 136) --- in fair agreement with the wavelet results.
The best-fit scintillation pattern parameters are reported in Table \ref{tab:amp}. The 
observed annual modulation cycle is compatible with the presence of a scattering
screen at a distance between 180\,pc and 270\,pc. The scintillation pattern is slightly anisotropic,
with an anisotropy degree around 2 and an anisotropy angle around 90 deg. Were we to assume that
the scattering screen is associated with the Ursa Major cloud complex, the estimated
distance would be compatible with the value of 240\,pc adopted by Heithausen et al. 
(\cite{Heithausen1993}), but less consistent with the distance of only 100--120\,pc
provided by Penprase (\cite{Penprase1993}) for the molecular clouds MBM 29--MBM 31 
in the complex. A distance to the screen of between 120 and 290\,pc can also be deduced 
from the Hipparcos catalog (Perryman \& ESA, \cite{Perryman1997}; van Leeuwen, 
\cite{vanLeeuwen2007}), by looking at the parallaxes of the stellar objects within a 
two-degree circle around S4~0954+658.

\section{Discussion}
\label{Discussion}

\begin{figure}[tbp]
   \centering
   \includegraphics[width=0.96\columnwidth]{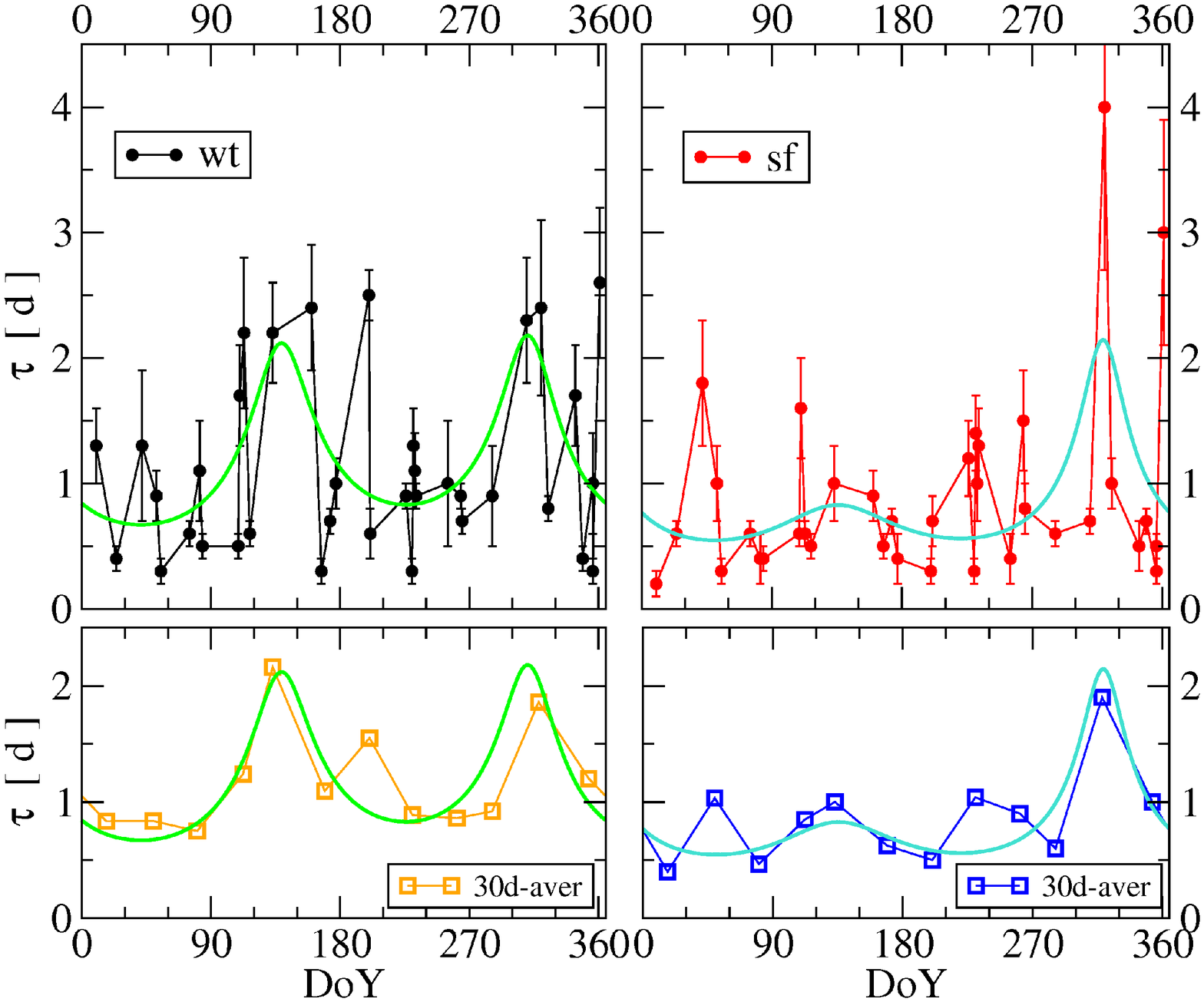}
      \caption{Annual modulation plots based upon the wavelet results (left panels),
      and the structure function results (right panels). The upper panels show all
      the estimated timescales, while lower panels show monthly averages. The 
      green and turquoise lines show the annual modulation patterns which best fit 
      the data.
              }
              \label{fig:am1}

   \centering
   \includegraphics[width=0.96\columnwidth]{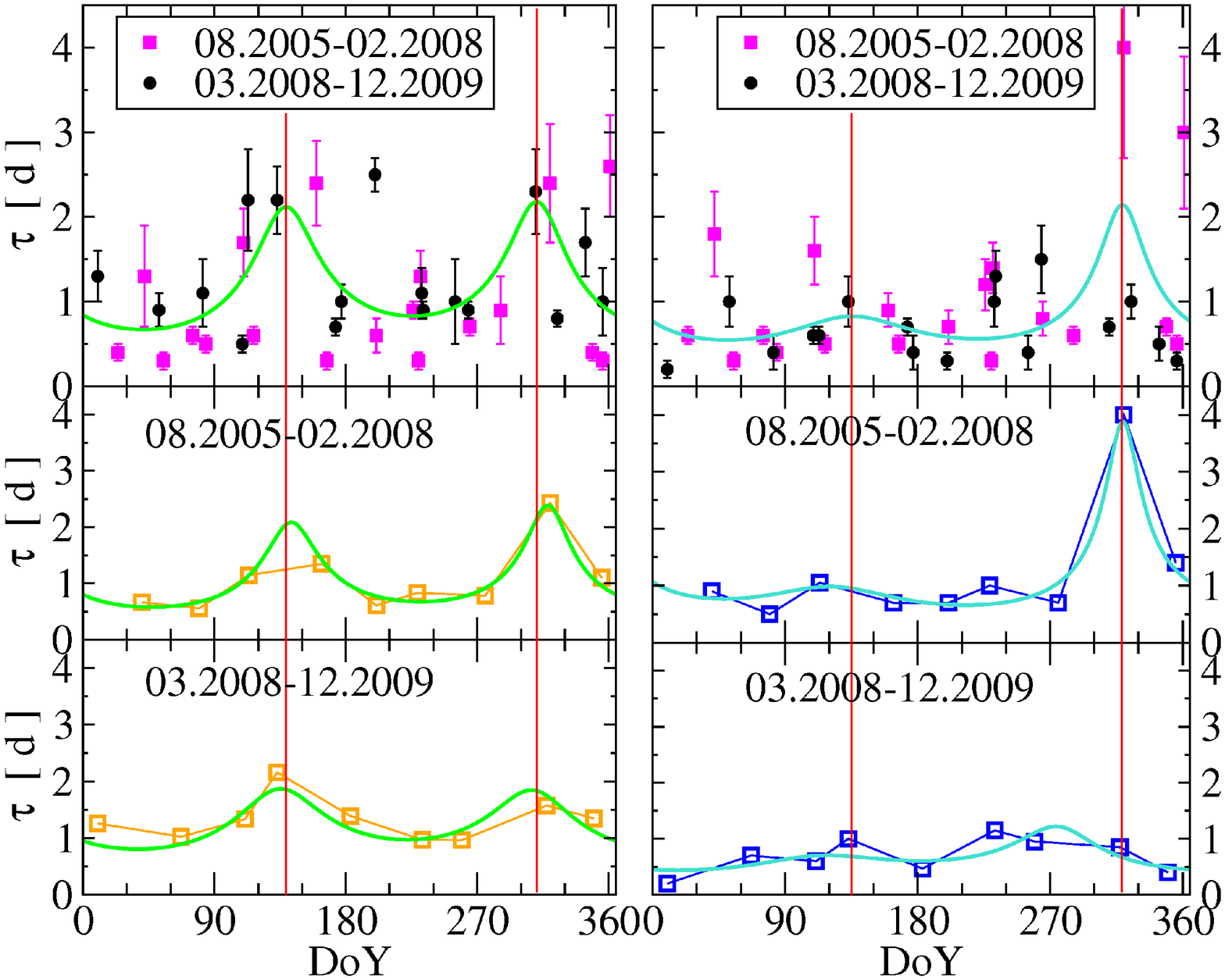}
      \caption{Annual modulation plots, as in Fig. \ref{fig:am1}; in the upper panels, the 
      timescales obtained for the epochs between August 2005 and February 2008 are 
      highlighted in magenta squares. In the middle and lower panels are
      displayed the annual modulation plots (green and turquoise lines) and the 40-day 
      timescales averages for the August 2005 -- February 2008 and the 
      March 2008 -- December 2009 epochs. The red vertical lines indicate the peaks in
      the annual modulation patterns obtained from the complete set of timescales.
              }
              \label{fig:am2}
\end{figure}

The existence of a break in the variability characteristics of the source,
described in Section \ref{Results}, raises the question of whether this break affects 
the search for an annual variation in the characteristic timescales. The 
consequences of the March 2008 break can be evaluated by separately analyzing 
the variability timescales before and 
after it occurs. The corresponding annual modulation plots and scintillation pattern
parameters are presented in Fig. \ref{fig:am2} and Table \ref{tab:amp} for the wavelet and 
structure functions separately. In the upper panels 
of Fig. \ref{fig:am2}, the August 2005 -- February 2008 timescales are shown as magenta 
squares, while the March 2008 -- December 2009 timescales are shown 
as black circles. In the middle and lower panels, we show 40-day 
averages of the timescales during the pre-outburst and the outburst phases, respectively. 
In all plots, the green and turquoise lines display the best annual modulation fits to the 
data. Apparently, the yearly cycles found before and after March 2008 in the 
wavelet-based results are quite similar, with slow-down phases at DoY 144 and 320 
for the former and at DoY 137 and 309 for the latter. In the case of 
the structure function results, instead, the difference in the cycles detected in the two timespans 
is significant. To understand the nature of the change in the variability characteristics of 
S4~0954+65 between the pre-outburst and the outburst phases, it is essential to find the reason 
why the wavelet and the structure function results are affected differently.

\subsection{Multiple timescales in the light curves after February 2008}

The general increase in the wavelet timescales after February 2008 could be explained 
in terms of a general slow-down in the variability of the source. 
This, however, could not account for the systematic discrepancies between the results
of the two analysis methods, highlighted in Fig.\,\ref{fig:ts_comp}. 
These discrepancies are most likely the consequence of the appearance, starting from 
February 2008, of multiple variability contributions in the light curves, of different 
timescales and comparable strengths. The wavelet and structure function plots in the right panels of 
Fig. \ref{fig:tam}, which show the results for one of the light curves collected during the 
outburst phase, confirm this interpretation. The wavelet plot shows two bumps, corresponding 
to timescales of 2.2 and 0.5 days, which indicate that there are two distinct variability
contributions; the two plateaus in the structure function plot, which have similar timescales, can be 
explained in the same way.

The discrepancies between the estimated timescales reveal the differences between the 
two analysis methods.
The wavelet timescales are generally longer than the structure function ones, suggesting
that the slowest variability contributions are the most powerful. For the structure 
function, it was our choice to identify the characteristic timescale with the one corresponding 
to the first significant plateau of the function. This rigid approach, which was forced by 
the necessity to minimize the arbitrariness of the definition of the characteristic timescales, 
should not interfere with the detection of possible annual modulation cycles. Assuming that a 
fast and a slow timescales are both detected in the majority of the 
light curves --- which is the case for the observing sessions during the outburst phase --- 
and that they are caused by the same scattering screen, they should follow the same annual 
modulation cycle. Therefore, theoretically, this can be identified by systematically looking at 
the fastest variability component --- which is the one that can be estimated with the highest 
accuracy, given the increase in the uncertainty with lengthening timescale. In practice, however,
the situation is more complicated; the large difference in the annual modulation plots of the
structure function timescales before and after March 2008  
demonstrates that a proper interpretation of the results cannot be achieved without 
simultaneously taking into account the information provided by both of the time 
analysis methods.

\subsection{The nature of the different variability components}

We present below two possible models for explaining the annual modulation
plots of the outburst phase of S4~0954+658 (see Fig. \ref{fig:am2}, lower panels). 
Both assume that a source of variability is the propagation effects
caused by a scattering screen between the source and the observer, and that a reasonable 
estimate of its main parameters is provided by the pre-outburst wavelet results (see 
Tab. \ref{tab:amp} and Fig. \ref{fig:am2}, middle-left panel). These assumptions are based 
on the existence of an annual cycle in the pre-outburst timescales and the detection of a similar 
cycle in the wavelet results during the outburst phase of the source. 

\begin{figure}[t!]
   \centering
   \includegraphics[width=0.96\columnwidth]{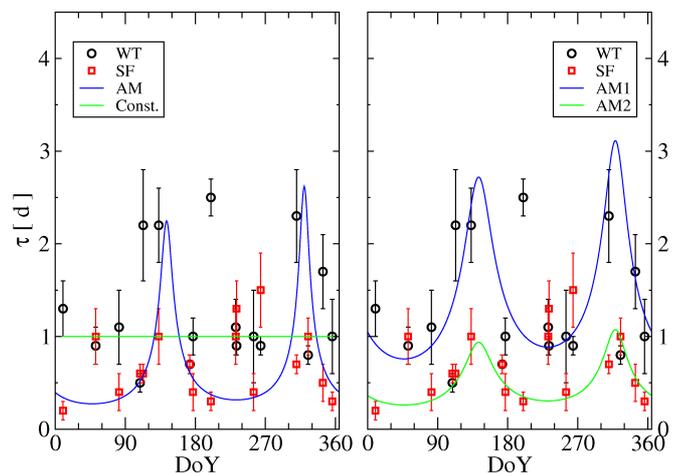}
      \caption{The wavelet (black dots) and structure function (red squares) timescales 
      calculated during the outburst phase of S4~0954+658 are indicative of
      multiple variability contributions. In the left panel, the variability is interpreted as the
      superposition of a source-extrinsic effect (blue line) and a source-intrinsic process (green 
      line). In the right panel, both the variability contributions are attributed to ISS and the two 
      annual modulation patterns have the same scattering screen parameters, the only 
      difference being the size of the scintillating components (90 $\mu$as for the blue line, 
      30 $\mu$as for the green line).
       }
              \label{fig:am-cost}
\end{figure}

In the first model, we hypothesize that the second contribution to the variability is 
provided by a source-intrinsic mechanism; the corresponding variability timescale 
should show no annual modulation effect. 
The frequent detection of timescales of about one day in the wavelet and structure function 
results suggests that the source-intrinsic process could be described in terms of a 
variability contribution of constant timescale. 
The present scenario is represented in Fig. \ref{fig:am-cost}, left panel. The wavelet 
and structure function results are plotted along with a one-day constant value (green
line)
and an annual modulation pattern (blue 
line), which assumes an intrinsic angular size of the scintillating component of
50 $\mu$as. To improve the fit to the data, the anisotropy degree of the scattering 
pattern was increased from three to seven. The most reasonable way of explaining this change is to hypothesize that the origin of the anisotropy is the emitting 
component of the source; structural changes in this component would also explain
the observed break in the variability characteristics of S4~0954+658. 

In the second model, both the variability contributions are caused by ISS through 
the same scattering screen, parameterized according to the fit to the pre-outburst 
wavelet results. The corresponding annual modulation curves, plotted in the right
panel of Fig. \ref{fig:am-cost}, assume intrinsic angular sizes of the scintillating 
components of 90 (blue line) and 30 $\mu$as (green line). Following Marscher et 
al. (\cite{Marscher1979}), and considering an amplitude of the IDV of $\sim 0.1$ Jy, 
from the size of the smaller component we can calculate its brightness 
temperature. This turns out to be higher than $10^{12}$ K; in order to avoid a
violation of the inverse-Compton limit, we have to assume that the emitting 
component is moving relativistically with a Doppler boosting factor $\ge 3$.

The characteristic timescales estimated during the outburst phase of S4~0954+658
seem compatible with both models. The source-intrinsic scenario provides a simple
explanation of the 11 occurrences of timescales of about one day detected in the 17 
observing sessions, but its fit to the wavelet timescales is worse than the one obtained
through the fully source-extrinsic scenario. None of the models explains 
the long timescale detected by the wavelet method in July 2008.

The limited amount of information provided by the variability analysis does not allow us to 
reach any final conclusion. Whatever the origin of the additional variability component,
its appearance is in all cases remarkable, especially when
associated with the outburst phase of the source. It seems unlikely that the simultaneity 
of the two events is accidental. Since the outburst phases of blazars are often explained
by changes in the structure of the sources, such as the emission of new emitting 
components (see, e.g., Mutel et al. \cite{Mutel1990}), and since the size and luminosity of 
the emitting regions are supposed to play a fundamental role in IDV, the existence of
a correlation between outburst phases and IDV characteristics of a source appears
pretty reasonable. A possible connection between changes in the variability patterns 
of IDV sources and variations in their structures was reported by 
Fuhrmann et al. (\cite{Fuhrmann2002}) and Macquart \& de Bruyn (\cite{Macquart2007}), and
our findings seem to confirm this picture. We note that many AGNs show intrinsic structural variability on timescales
of months to years. These varying source structures complicate the search for annual modulation patterns, which may
be detectable in AGNs only during sufficiently long periods of quiescence.

\section{Summary}

We have reported on the evolution of the intraday variability characteristics of S4~0954+65
during the period from August 2005 to December 2009. For this source, 41 IDV
observing sessions have been collected at the Urumqi Astronomical Observatory at a 
frequency of 4.8 GHz; 38 of these sessions were long enough to provide a reliable 
estimation of the variability characteristics.

The data analysis has been performed using standard tools of statistics and time series 
analysis, such as the variability amplitude, the chi-square test for evaluating the probability of 
a constant flux density and the first-order structure function. A new method for the estimation
of the characteristic timescale --- a modified wavelet analysis based on the Ricker mother 
wavelet --- has been developed and successfully applied to the data. The results provided by 
this new algorithm show a high degree of correlation with the ones obtained using a structure 
function analysis. We collated a comparison data set of more than 120 light curves, comprising the light curves of all the main targets of the monitoring program except S4~0954+65; for this data set, a linear regression between wavelet and structure function timescales recovers a correlation
coefficient of 0.86 and a slope of 1.01, which demonstrates the consistency between the definitions 
of timescales in the two methods. The wavelet estimation of the timescales offers the
important advantages of providing an unambiguous definition of the variability timescale and
the possibility of full automation. 

The combined use of a wavelet and structure function analysis proved to be a powerful tool for
tracing the evolution of the IDV timescales of S4~0954+65. Both methods reveal that there is an annual modulation cycle in the variability timescales, suggesting an essential
contribution of propagation effects to the detected variability. The seasonal cycle is not as strong
as the ones detected in fast scintillators, such as J1819+3845 (Dennett-Thorpe \& de Bruyn 
\cite{Dennett2000}), or in the type-II source J1128+5925 (Gab\'anyi et al. \cite{Gabanyi2007}),
but the agreement between the cycles observed in the wavelet timescales before and after February 2008,
with slow-down phases around DoY 140 and 310, and the confirmation of a slow-down phase around 
DoY 320 in the structure function results are strong arguments in favor of a source-extrinsic origin
of the variability.

The comparison of the results from the two methods contributed to the discovery of a substantial 
change in the IDV 
characteristics of the source, occurring between February and March 2008 at the beginning 
of a strong outburst phase. The signature of this change is a sudden lengthening in the average
timescales detected by the wavelet-based analysis, an increase in the variability amplitude
measured through the structure function, and a dramatic weakening in the correlation among the 
results provided by the two methods. The variability detected in the light curves is likely the
result of the superposition of two independent contributions. The data seem to favor
a source-extrinsic origin for both the contributions, but the results cannot be unambiguously 
interpreted. Nevertheless, 
the clear correlation between the long-term flux density variations in S4~0954+65 --- likely 
associated with changes in the source structure --- and its IDV features shows that the intrinsic 
properties of the source exert a significant influence on the characteristics of the variability 
on the shortest timescales.

\begin{acknowledgements}
We would like to thank Peter M\"uller, who provided the Python-based software 
package TOOLBOX for the processing of the Urumqi raw data. N.M. is funded by 
an ASI fellowship under contract number I/005/11/0. K. \'E. G. was 
supported by the Hungarian Scientific Research Fund (OTKA, K72515). This paper 
made use of data obtained with the 25m Urumqi telescope of the Xinjiang Astronomical 
Observatory  of the Chinese Academy of Sciences (CAS). Liu X. is supported by the 
National Natural Science Foundation of China under grant No.11073036
and the 973 Program of China (2009CB824800).
\end{acknowledgements}

\begin{appendix}
\section{Wavelet analysis}

The wavelet-based analysis we used to estimate the timescales in
our datasets is based on the Ricker mother wavelet  

\begin{equation}
\Psi(t)=C\,\big(1-t^2\big)\,exp\big(-\frac{t^2}{2}\big),
\end{equation}
where $C$ is a normalization factor and $t$ is the time. Given the
scale $s$ and the time-shift $l$, we define the wavelet 

\begin{equation}
\Psi_{s,\,l}(t)=\sum_{i}^{N}\,\Big(1-\frac{(t_\mathrm{i}-l)^2}{s^2}\Big)\,exp\Big(-\frac{(t_\mathrm{i}-l)^2}{2\,s^2}\Big)
\end{equation}
and the wavelet transform

\begin{equation}
\psi(s, l)=\sum_{i}^{N}\,(S_\mathrm{i}-<S>)\,\Big(1-\frac{(t_\mathrm{i}-l)^2}{s^2}\Big)\,exp\Big(-\frac{(t_\mathrm{i}-l)^2}{2\,s^2}\Big).
\label{eq:wav}
\end{equation}
For each variability curve, we calculated $\psi(s, l)$ on a range of scales
$s$, from twice the average sampling to half the total length of the
curve, with $l$ varying between $(t_0+s/2)$ and $(t_N-s/2)$; we defined the
variability timescale as $\sqrt{3}\,s_M$, where $s_M$ is the scale for which
$\psi(s, l)$ is maximum, noting that, given a scale $s$, $\sqrt{3}\,s$ is the
time lag between the minimum and the maximum of $\Psi_{s,\,l}(t)$.

The results provided by the method described above are not satisfactory. 
To allow a proper comparison between the wavelet transforms,
$\sum_i^N \Psi_{s,\,l}(t_i)$ should be close to zero for each set of
parameters $(s, l)$. This condition is often unfulfilled, because of the
uneven sampling of the variability curves and their limited duration. 
To correct for this, we introduced a modified wavelet transform

\begin{equation}
\phi(s, l)=C_{s,\,l}\,\psi(s, l),
\label{eq:modwav}
\end{equation}
where $C_{s,\,l}$ is defined as

\begin{equation}
C_{s,\,l}=\sqrt{\sum_{i}^{N}\Big(\big(1-\frac{(t_\mathrm{i}-l)^2}{s^2}\big)\,exp\big(-\frac{(t_\mathrm{i}-l)^2}{2\,s^2}\big)\Big)^2}.
\label{eq:cls}
\end{equation}
As for the standard wavelet analysis, the variability timescale
is defined as $\sqrt{3}\,s_M$, where $s_M$ is the scale that corresponds to the maximum value of
$\phi_{s,\,l}$. The standard wavelet analysis is generally more sensitive to slow variability components
than to fast ones, often detecting unreasonably long characteristic timescales. The results 
provided by the modified wavelet, instead, appear to be much more reliable.

The modified wavelet transform has a strong advantage over the structure 
function analysis in dealing with light curves with variability on 
multiple timescales. While the identification of the characteristic timescale 
in the structure function requires a subjective interpretation 
of which timescale is the most predominant, the wavelet analysis always provides 
a unique value, because the set $(s, l)$ that maximizes $\phi_{s,\,l}$ is always 
univocally defined. This completely removes the ambiguity when interpreting
the analysis results --- regardless of who performs the analysis, the same 
timescale is found. It also implies that the wavelet analysis can be fully automated, which 
can be very useful in dealing with large datasets. On the other hand, one 
should always keep in mind that the timescale obtained by the wavelet 
method, strictly speaking, does not provide an estimation of the \emph{characteristic} 
variability timescale. It refers, instead, to the strongest variability feature in the light 
curve. When the variability timescale in a light curve is
much shorter than the duration of the observations, the structure
function should  be regarded as a more suitable tool for the data analysis. 

We note that the intraday variability of blazars --- such as
S4~0954+65 --- at radio frequencies can be generally described as red
noise. The amplitude of the variability is stronger on longer
timescales, which implies that the variability timescale is typically
on the same order of magnitude as the duration of the observations. In
these conditions, the identification of the characteristic timescales
with the strongest variability feature in the light curves appears
quite reasonable. This conclusion is convincingly supported by the
high degree of correlation between the structure function and the
wavelet results (see Section \ref{Results}). 

The error in the timescale estimates provided by the modified
wavelet transform can be ascribed to two main contributions. The first
consists in the combined effect of the uneven sampling and the
uncertainty in the flux density measurements; the second has to do with
the limited duration of the observations, which causes a gradual
decrease in the reliability of the wavelet transforms on larger scales
$s$. Both effects are reflected in the width of the peak $\psi(s_M,
l)$. Wider peaks are indicative of a low contrast among the strengths of the
variability on different scales $s$, hence larger errors in the
estimation of the timescale. We performed tests with synthetic
light curves of different durations, sampled both evenly and unevenly.
We found that a reasonable evaluation of the error in the timescales is 
given by $(s_2-s_1)/2$,  where $s_2$ and $s_1$ are respectively the 
maximum and the minimum scales for which $\psi(s, l)$ equals 
$0.95 \cdot \psi(s_M, l)$.

\end{appendix}


\begin{thebibliography}{.99}

\bibitem[1992]{Bevington1992} Bevington, P.~R., \& Robinson, D.~K.\ 1992, New York: McGraw-Hill, |c1992, 2nd ed.,  

\bibitem[2003]{Bignall2003} Bignall, H.~E., 
Jauncey, D.~L., Lovell, J.~E.~J., et al.\ 2003, \apj, 585, 653 

\bibitem[2006]{Bignall2006} Bignall, H.~E., 
Macquart, J.-P., Jauncey, D.~L., et al.\ 2006, \apj, 652, 1050 

\bibitem[2002]{Cimo2002} Cim{\`o}, G., Beckert, 
T., Krichbaum, T.~P., et al.\ 2002, \pasa, 19, 10 

\bibitem[2000]{Dennett2000} Dennett-Thorpe, J., \& de Bruyn, A.~G.\ 2000, \apjl, 529, L65 

\bibitem[1987]{Fiedler1987} Fiedler, R.~L., Dennison, B., Johnston, K.~J., \& Hewish, A.\ 1987, \nat, 326, 675

\bibitem[1994]{Fiedler1994} Fiedler, R., Dennison, B., Johnston, K.~J., et al. 1994, \apj, 430, 581 

\bibitem[2002]{Fuhrmann2002} Fuhrmann, L., 
Krichbaum, T.~P., Cim{\`o}, G., et al.\ 2002, \pasa, 19, 64

\bibitem[2004]{Fuhrmann2004} Fuhrmann, L. 2004, PhD thesis, Bonn University, Germany

\bibitem[2007]{Gabanyi2007} Gab{\'a}nyi, K.~{\'E}., Marchili, N., Krichbaum, T.~P., et al.\ 2007, \aap, 470, 83 

\bibitem[1984]{Heeschen1984} Heeschen, D.~S.\ 1984, \aj, 89, 1111 

\bibitem[1987]{Heeschen1987} Heeschen, D.~S., 
Krichbaum, T., Schalinski, C.~J., \& Witzel, A.\ 1987, \aj, 94, 1493 

\bibitem[1993]{Heithausen1993} Heithausen, A., Stacy, J.~G., de Vries, 
H.~W., Mebold, U., \& Thaddeus, P.\ 1993, \aap, 268, 265

\bibitem[1997]{Kedziora1997} 
Kedziora-Chudczer, L., Jauncey, D.~L., Wieringa, M.~H., et al.\ 1997, 
\apjl, 490, L9 

\bibitem[1969]{Kellermann1969} Kellermann, K.~I., \& Pauliny-Toth, I.~I.~K.\ 1969, \apjl, 155, L71 

\bibitem[1971]{Kinman1971} Kinman, T.~D., \& Conklin, E.~K.\ 1971, \aplett, 9, 147 

\bibitem[2003]{Kraus2003} Kraus, A., Krichbaum, T.~P., Wegner, R., et al.\ 2003, \aap, 401, 161 

\bibitem[2003]{Lovell2003} Lovell, J.~E.~J., 
Jauncey, D.~L., Bignall, H.~E., et al.\ 2003, \aj, 126, 1699 

\bibitem[2008]{Lovell2008} Lovell, J.~E.~J., 
Rickett, B.~J., Macquart, J.-P., et al.\ 2008, \apj, 689, 108 

\bibitem[2007]{Macquart2007} Macquart, J.-P., \& de Bruyn, A.~G.\ 2007, \mnras, 380, L20

\bibitem[2009]{MarchiliTh} Marchili, N. 2009, PhD thesis, Bonn University, Germany

\bibitem[2010]{Marchili2010} Marchili, N., Mart{\'{\i}}-Vidal, I., Brunthaler, A., et al.\ 2010, \aap, 509, A47

\bibitem[2011]{Marchili2011}  Marchili, N., Krichbaum, T.~P., Liu, X., et al.\ 2011, \aap, 530, A129 

\bibitem[1979]{Marscher1979} Marscher, A.~P., 
Marshall, F.~E., Mushotzky, R.~F., et al.\ 1979, \apj, 233, 498

\bibitem[1990]{Mutel1990} Mutel, R.~L., Phillips, 
R.~B., Su, B., \& Bucciferro, R.~R.\ 1990, \apj, 352, 81 

\bibitem[1992]{Narayan1992}  Narayan, R.\ 1992, Royal 
Society of London Philosophical Transactions Series A, 341, 151 

\bibitem[2001]{Qian2001} Qian, S.-J., \& Zhang, X.-Z.\ 2001, \cjaa, 1, 133 

\bibitem[1991]{Quirrenbach1991} Quirrenbach, A., 
Witzel, A., Wagner, S., et al.\ 1991, \apjl, 372, L71 

\bibitem[1992]{Quirrenbach1992} Quirrenbach, A., Witzel, A., Kirchbaum, T.~P., et al.\ 1992, \aap, 258, 279 

\bibitem[2000]{Quirrenbach2000} Quirrenbach, A., Kraus, A., Witzel, A., et al.\ 2000, \aaps, 141, 221 

\bibitem[1993]{Penprase1993} Penprase, B.~E.\ 1993, \apjs, 88, 433 

\bibitem[1997]{Perryman1997} Perryman, M.~A.~C., \& ESA 1997, ESA Special Publication, 1200,

\bibitem[1984]{Rickett1984} Rickett, B.~J., Coles, W.~A., \& Bourgois, G.\ 1984, \aap, 134, 390 

\bibitem[2001]{Rickett2001} Rickett, B.~J., Witzel, 
A., Kraus, A., Krichbaum, T.~P., \& Qian, S.~J.\ 2001, \apjl, 550, L11 

\bibitem[1978]{Shapirovskaya1978} Shapirovskaya, N.~Y.\ 
1978, \sovast, 22, 544 

\bibitem[1985]{Simonetti1985} Simonetti, J.~H., 
Cordes, J.~M., \& Heeschen, D.~S.\ 1985, \apj, 296, 46 

\bibitem[2006]{Sun2006} Sun, X.~H., Reich, W., Han, J.~L., Reich, P., \& Wielebinski, R.\ 2006, \aap, 447, 937 

\bibitem[2007]{vanLeeuwen2007} van Leeuwen, F.\ 2007, \aap, 474, 653

\bibitem[1990]{Wagner1990} Wagner, S., Sanchez-Pons, F., Quirrenbach, A., \& Witzel, A.\ 1990, \aap, 235, L1 

\bibitem[1993]{Wagner1993} Wagner, S.~J., Witzel, A., Krichbaum, T.~P., et al.\ 1993, \aap, 271, 344 

\bibitem[1971]{Wills1971} Wills, B.~J.\ 1971, \apj, 169, 
221 

\bibitem[1986]{Witzel1986} Witzel, A., Heeschen, D.~S., Schalinski,
  C., \& Krichbaum, T.~P.\ 1986, Mitteilungen der Astronomischen
  Gesellschaft Hamburg, 65, 239  


\end{thebibliography}
\end{document}